\documentclass[a4paper]{aa}
\usepackage{graphicx}
\usepackage{natbib}
\usepackage{babel}
\usepackage{hyperref}
\usepackage{color}
\usepackage{aalongtable}
\usepackage{multicol}

\def\Teff{$T_{\rm eff}$~}
\def\logg{$\log~g$~}

\def\vsini{$V$~sin~$i$~}
\def\Vt{$V_{\rm t}$~}
\def\Vm{$V_{\rm macro}$~}

\def\ctwo{C$_2$~}

\def\kmps{kms$^{\rm -1}$~}

\begin{document}

\title{Masses, Oxygen and Carbon abundances in CHEPS dwarf stars}

\author{Y. V. Pavlenko$^{1,2,3}$, B. M. Kaminsky$^2$, J. S. Jenkins$^{4,5}$, O. M. Ivanyuk$^{2}$, H. R. A. Jones$^{1}$, Y. P. Lyubchik$^{2}$}

\institute{Centre for Astrophysics Research, University of Hertfordshire, College Lane, Hatfield, Hertfordshire AL10 9AB, UK \and Main Astronomical Observatory, NAS of Ukraine, Akademika Zabolotnoho, 27, Kyiv, 03143, Ukraine \and Nicolaus Copernicus Astronomical Center, PAS, Rabianska, 8, 87-100 Toru\' n, Poland \and Departamento de Astronom\'ia, Universidad de Chile, Casilla 36-D, Santiago, Chile \and Centro de Astrof\'isica y Tecnolog\'ias Afines (CATA), Casilla 36-D, Santiago, Chile\\}

\offprints{email2yp@gmail.com} \mail{email2yp@gmail.com} \date{}

\authorrunning{Y. V.Pavlenko et al.} \titlerunning{Carbon and Oxygen in 107 dwarf stars}

\abstract{
We report the results from the determination of stellar masses, carbon and oxygen abundances in the atmospheres of 107 stars from the CHEPS program. Our stars are drawn from a population with a significantly super-solar metallicity. At least 10 of these stars are known to host orbiting planets.
}{
In this work, we set out to understand the behavior of carbon and oxygen abundance 
in stars with different spectral classes, metallicities and \vsini , within the metal-rich stellar population.
}{
Masses of these stars were determined using the data from Gaia DR2 release. The oxygen and carbon abundances were determined 
by fitting the absorption lines. 
Oxygen abundances were determined by fits to the 6300.304 \AA~ O {\sc I} line, and for the determination of the carbon abundances we used 3 lines of the C {\sc I} atom and 12 lines of C$_2$ molecule.
}{
We determine masses and abundances of 107 CHEPS stars.
There is no evidence that the [C/O] ratio depends on V sin i or the mass of the star, within our constrained range of masses, i.e. $0.82 < M_*/M_{\odot} < 1.5$ and metallicities  $-0.27 < [Fe/H] < + 0.39$ and we confirm that metal-rich dwarf stars with planets are more carbon-rich in comparison with non-planet host stars, with a statistical significance of 96\%.
}{
We find tentative evidence that there is a slight offset to lower abundance and a greater dispersion in oxygen abundances relative to carbon, and interpret this as potentially arising from the production of the oxygen being more effective at more metal-poor epochs. We also find evidence that for lower mass star’s the angular momentum loss in star’s with planets as measured by V sin i is steeper than star’s without planets. In general, we find that the fast rotators (\vsini $>$ 5 \kmps ) are massive stars. 
}


\maketitle

\section{Introduction}
\label{_intro}

In this work we pay special attention to the determination of the carbon and oxygen abundance of a sample of metal-rich dwarf stars, mainly of spectral class G. Understanding the chemical make-up of G-stars is fundamental to our understanding of star formation and stellar evolution. In many ways, G-dwarfs are key objects to enhance our understanding of Galactic evolution. For instance, observations indicate that there are too few metal deficient G dwarfs ('G-dwarfs problem') with respect to that which could be expected from simple models of chemical evolution in the Galaxy (e.g. \citealp{sear72}, \citealp{hayw01}), as well as other bulge dominated or disc dominated galaxies (\citealp{wort96}, see more details in \citealp{caim11}).

Carbon and oxygen were synthesized in the post Big Bang  epoch. However, they were formed by different processes. Carbon acts as a primary catalyst for the nuclear H-burning via the CNO-cycle. It was produced by synthesis 
in stellar interiors, dredged up from their cores, and then iteratively placed into the interstellar medium by powerful winds, driven by radiation  pressure from massive stars (e.g. \citealp{gust99}). The element contributes significantly to the stellar interior and atmospheric opacity. Carbon also plays an important role in dust formation processes in the interstellar medium.

Oxygen, on the other hand, is the third most common element overall, and along with its isotopes, provides key
tracers of the formation and evolution of planets, stars and galaxies and as such it is one of the most important elements in all of astronomy. 
Interestingly, its abundance was often disputed for a long time, even in the case of the solar abundance. 
In recent years the solar oxygen content has changed from log N(O)=-3.11$\pm$0.04 in 
\cite{ande89} to -3.33 in \cite{aspl05} or -3.35 \cite{aspl09}.

The studies of C and O abundances have been intensified in recent years. In an attempt to discover any differences between stars 
with planets (hereafter SWP) and non-SWP, \cite{delg10} performed a detailed study of C and O (as well as Mg and Si) abundances for a sample of 100 and 270 stars with and without known giant planets with effective temperatures
between 5100 K and 6500 K. Interestingly, these authors, together with \cite{escu04, escu06}, claimed the 
presence in the Galaxy of a large number
of carbon-rich dwarfs, i.e. dwarfs with C/O>1, see fig.1 in \cite{fort12}. However,
as was noted by \cite{fort12}, other authors (see \cite{deko95, down04, cove08}) found a much lower number of carbon rich dwarfs in our Galaxy. 

These recent studies of carbon and oxygen were triggered by some theoretical predictions that C/O (and
Mg/Si) are the most important elemental ratios in determining the mineralogy of terrestrial planets,
and they can give us information about the composition of these planets. Namely, the C/O ratio controls the
distribution of Si among carbide and oxide species, while Mg/Si gives information about the silicate
mineralogy \citep{bond10a, bond10b}. \cite{delg10} did not observe any notable difference 
between the abundances of SWP and non-SWP stars. However, the authors noted that the investigated sample of stars was 
not large enough to discard a possible effect due to the presence of
planets. Nevertheless, \cite{suar18} highlight a diversity of mineralogical ratios that reveal the different kinds of 
planetary systems that can be formed, most of them dissimilar to our solar system.  Different 
values of the Mg/Si and C/O ratios can determine different compositions of planets formed. 
They found that 100\% 
of their sample of stars with planets present
C/O < 0.8. 86\% of stars with high-mass 
companions present 0.8 > C/O > 0.4, while 14\% present C/O values lower than 0.4. 
Planet hosts with low-mass companions present C/O and Mg/Si ratios similar to those found in the Sun, 
whereas stars with high-mass companions have lower C/O.

In a related paper, \cite{suar17} studied carbon solar-type stars from the CH band at 4300 \AA.
They confirm two different slope trends for [C/Fe] vs. [Fe/H], realizing that the behavior changes 
for stars with metallicities above and below solar, and they obtained abundances and distributions
that show that SWP are more carbon rich when compared to single stars, a signature caused by the known
metal-rich nature of SWP. They found no different behavior when separating the stars by the mass of the
planetary companion. Furthermore, they claimed a flat distribution of the [C/Fe] ratio for all planetary masses, that
apparently excludes any clear connection between the [C/Fe] abundance ratio and planetary mass.

The layout of the manuscript is as follows,
in Section \ref{_obs1} we provide information about the stars of our sample and
the observed spectra,
in Section \ref{_pro} we discuss the details of our procedure of carbon and oxygen abundance determination,
Sections \ref{_res} contains the description of our measured carbon and oxygen abundance
results. In Section \ref{_dis} we summarise our findings. In the Appendix we provide some 
information about used line lists, examples of fits to observed profile of O I line. Here we show 
histograms of carbon distributions in the SWP and non-SWP stars, plots of [C/Fe] and [O/Fe] vs. age
of stars of our sample.

\section{Observations and data acquisition} 
\label{_obs1}

We used the observed spectra obtained in the framework of the Calan-Hertfordshire Extrasolar Planet Search (CHEPS) programme \citep{jenk09}. The programme was proposed to monitor samples of metal-rich dwarf and subgiant stars selected from Hipparcos, with $V$-band magnitudes in the range 7.5 to 9.5 in the southern hemisphere, in order to search for planets that could help improve the existing statistics for planets orbiting such stars.

The selection criteria for CHEPS was based on selecting inactive (log$R'_{\rm{HK}}\le$-4.5\,dex) and metal-rich ([Fe/H]$\ge$+0.1\,dex) stars by the analysis of high-resolution FEROS spectra \citep{jenk08,jenk11,murgas13} to ensure the most radial velocity stable targets, and to make use of the known increase in the fraction of planet-host stars with increasing metallicity, mentioned above. Furthermore, SIMBAD\footnote{simbad.u-strasb.fr/simbad} does not indicate our stars are members of binary systems. However, we have discovered a number of low-mass binary companions as part of the CHEPS programme (e.g. \citealp{jenk09}; \citealp{pant18}).

All stars in our work were observed with the HARPS spectrograph \citep{mayor03} at a resolving power of 115,000, and since the spectra were taken as part of the CHEPS programme, whose primary goal is the detection of small planets orbiting these stars, the S/N of the spectra are all over 100 at a wavelength of 6000 \AA. The 107 stars in this work are primary targets for CHEPS. However, there are additional targets that have been observed with CORALIE and MIKE that we have not included in this work to maintain the homogeneity of our analysis, specifically these other instruments operate at significantly lower resolution than HARPS.

Thus far the CHEPS project has discovered 15 planets \citep[see][]{jenk17} and a number of brown dwarfs and binary companions \citep{vine18}. 
The high-resolution and high-S/N of the 107 CHEPS spectra from HARPS allowed \cite{ivan17} and \cite{soto18} to study chemical abundances like Na, Mg, Al, Si, Ca, Ti, Cr, Mn, Fe, Ni, Cu and Zn in the atmospheres of metal-rich dwarfs. Gravities in the atmospheres of subgiants are lower in comparison to dwarfs, however, we carried out the same procedure for the lithium abundance determination for the stars of both groups \citep{pavl18}. Differential analysis of the results allows us to investigate the effects of, for example, gravity and effective temperature on the present stages of evolution of our stars.

\section{Procedure}
\label{_pro}

\subsection{Basic parameters and abundances for the stars}

We computed masses of our stars using the Gaia 2nd release \citep{gaia18} and the Mass-Luminosity empirical 
relationship of \cite{eker15}. These masses are shown in Table \ref{_ACOres}. In the following we mark the stars of larger masses by larger circles in our figures.

We used the stellar atmospheric parameters (effective temperatures \Teff, gravities \logg, microturbulent velocities \Vt, rotational velocities \vsini and abundances of Na, Mg, Al, Si, Ca, Ti, Cr, Mn, Fe, Ni, Cu, and Zn determined by \citet{ivan17}, who used our procedure of finding the best fit of the synthetic absorption line profiles to the observed spectra using the ABEL8 program \citep{pavl17}. We applied the model atmospheres computed by \cite{ivan17} using the SAM12 programme \citep{pavl03}. Model atmospheres and synthetic spectra were computed for the same set of input parameters. This became the basis for our abundance calculations that we detail below.
For the Sun we adopt the following parameters for the solar atmosphere \Teff/\logg/[Fe/H] = 5777/4.44/0.0, and we also adopt the solar abundances determined by \cite{ande89}.

In this paper we use 'Kurucz's' abundance scale in which $\sum$ N(X$_i$) = 1.0, where N(X$_i$) is the
relative number of the $i$th-element. Our abundances can be transferred into other popular abundance scales in which 
log N(H)=12.00 by log N(X)=log N(x)-12.04. We also use the classical definitions of abundances or abundance ratios
measured 'relative to the Sun': [X$_i$]=log N(X$_i$)-log N$_\odot$(X$_i$) and
[X/Y]=log N(X)-log N(Y)-(log N$_\odot$(X)-log N$_\odot$(Y)). In
this paper we adopt the \cite{ande89} abundances
as the reference ones. Generally speaking, we performed the abundance analysis in some sense 'relative to the Sun'. All abundances and abundance ratios can be reduced to any adopted solar abundance scale, for example \cite{aspl09}.
 
\subsection{Model atmospheres}

We used the 1D LTE model atmospheres computed by \cite{ivan17}. These model atmospheres were not recomputed in the process of the iterative determination of C and O, because the response of the model atmospheres to C and O variations is rather marginal.

\subsection{Line lists}

Different authors do not use the same line lists for the carbon
abundance determination. \cite{delg10} used C {\sc I} lines at 5380.3 and 5052.2 \AA, with 5380.3 \AA\ only used for stars with \Teff$<$5100 K. For oxygen, 
the forbidden lines of [O {\sc I}] at 6300.304 and 6363 Å were used. A detailed study
of these lines was carried out by \cite{bert15}.
\cite{peti11} used a C {\sc I} line for carbon at 
6587 \AA, and the [OI] line for oxygen at 6300.304 \AA.
The abundances were determined by the Spectroscopy Made Easy (SME)
code \citep{vale96} with Kurucz stellar atmospheres.
\cite{alex15} constructed a comprehensive model
atom for C {\sc I} using the most up-to-date atomic data 
available so far, to determine the carbon abundance of the
Sun and selected late-type stars with well-determined stellar 
parameters based on the LTE and NLTE line-formation for C {\sc I} in
the classical 1D model atmospheres and high-resolution observed spectra. The small sample of stars covers the -2.58$<$[Fe/H]$<$0.00 range. They also derive the carbon abundance from the
molecular CH and \ctwo lines and investigate the differences
between the atomic and molecular lines. This study shows rather 
marginal NLTE abundance corrections are required for atomic lines of C {\sc I}. 

For the determination of carbon abundance in the atmospheres of our stars we used the line list of the CCS (Carbon Contained Species) from the paper of \cite{alex15}. 
In total, the list of C {\sc I}, CH, \ctwo lines consists of 24 features. They are listed in the Table \ref{_Accs}. 
For simplicity we adopt that all carbon exists in the form of $^{12}$C atoms. 
The procedure of line selection and verification of the spectroscopic data are described in detail in their paper. Here we note that to
avoid possible problems with blending in the wings of the absorption lines, 
we fitted our synthetic spectra to the cores of
observed features, for which the problem is weaker, \citep[see][]{pavl17}. 

Oxygen abundances were determined from the fits of computed spectra to the observed feature at 6300.3 \AA. \cite{alle01} showed that the oxygen line 6300.3\,\AA\ forms a blend with a few other lines, with the Ni {\sc I} line at 6300.58 \AA\
dominating, see Table \ref{_AOIl} and Fig. \ref{_AoinSun}.

\subsection{C and O abundance analysis}

In the framework of our approach we assume that

-- all lines follow a Voigt profile;

-- blending was treated explicitly, and in our work we use the spectroscopic data of atomic lines from the VALD-3 database \citep{ryab15};
except for the case of some specifically selected lines, 
 
-- both the instrumental broadening and macroturbulent broadening may be described by Gaussian profiles.

We adopted that a macroturbulent velocity \Vm distribution in the atmospheres of our stars is similar to the case of the solar atmosphere.
For the case of the solar spectrum, the macroturbulent velocities mean that the measured widths of solar lines corresponds to a resolving power,
$R$=70K at 6700 \AA. This formal resolution is limited by the presence in the solar atmosphere of macroturbulent
motions with \Vm = 1.0-2.6 \kmps, see \cite{pavl12} and references therein. 
We used this value of $R$ for all our spectra. Generally speaking, \Vm varies with depth in the atmosphere, depending on
the physical state of the outer part of the convective envelope of a star. 
It is worth noting that in the case of slow rotators, the determination of \vsini and \Vm is a degenerate problem. On the other hand,
instrumental broadening, rotational broadening and broadening by macroturbulent velocities do not affect the 
integrated intensity of absorption lines. We adopt the \vsini values of our stars determined by \cite{ivan17}.
To obtain more accurate fits to the observed spectrum we used the \Vm parameter
to adjust our fits to the observed line profiles, by varying this velocity around the adopted value by a few \kmps.

Our simplified model does not consider 
changes of \Vt and \Vm with depth, differential rotation 
of stars, presence of spots, and the effects of magnetic activity, among others.
We believe that taking account of these effects should not significantly alter our results, at least as relates to the 
C and O abundance determination. The detailed modeling
of any of the listed effects we outline here requires very sophisticated analyses that is beyond the framework of our paper.

The effects of line blending were treated explicitly, whereby only well fitted parts of the observed line profile were used in our analysis. This approach allowed us to minimize effects of blending by lines of other atoms.

\subsection{Problems of oxygen abundance determination}
\label{_oproblems}

The strong absorption feature at 6300.304 \AA\ is formed by blends of a few lines, see Table \ref{_AOIl}. 
In the spectra of solar like stars the main 
contributors here are the lines of O {\sc I} and Ni {\sc I}, \citep[see][]{alle01}. Other lines
are less important, at least for the case of the Sun and similar stars. 
Following the approach developed by \cite{alle01} we changed the gf=1.832e-2 of the Ni {\sc I} line \citep{ryab15}
to gf=1.000e-3. 
With the solar nickel abundance log N(Ni)=-5.79 we obtained the solar 
oxygen abundance log N(O)=-3.11 \citep{ande89}, showing we can attain a good fit to the observed O {\sc I} profile, see Fig. \ref{_AoinSun}.

\begin{table*}
\caption{The oscillator strength $gf$ of O {\sc I} and Ni {\sc I} lines from different sources}
\begin{tabular}{|c|c|c|c|}
\hline
&&& \\
Element    &     O {\sc I}   &      Ni {\sc I}  &    References \\  
&&& \\
\hline
&&& \\
$\lambda$,\AA        & 6300.304  &   6300.336 & \\
&&& \\ \hline
&&& \\
          & 6.281E-13 &   1.832E-02 & VALD \cite{ryab15} \\
  $gf$     & 1.919E-10 &   4.898E-03 & 3D \cite{alle01}   \\
           & 1.517E-10 &   7.327E-03  &  1D \cite{bens04} \\
           & 1.919E-10 &   7.762E-03   &  1D \cite{bert15} \\
           & 1.919E-10 &   2.500E-03 & 1D, adopted in this paper \\
           &           &             &                \\
           \hline
\end{tabular}
\end{table*}

This simplifies the procedure of comparison of the observed and computed spectra, 
on the other hand, telluric O$_2$ lines in the observed spectrum move to new positions 
and for some stars one of these telluric lines moves onto the
O {\sc I} 6300.304 line. In Fig. \ref{_A6300} we provide the spectral ranges
around 6300 \AA\ for the spectra of three stars, in one case, i.e. HIP 29442 we get a strong absorption line at the 
position of 6300.304 \AA.
The relatively constant strength of these telluric O$_2$ lines (see http://www.astrosurf.com/buil/us/spe2/hresol4.htm)  
means that such cases are easily identified by eye, see Fig. \ref{_A6300}.

The use of another O {\sc I} line at 6363.776 \AA\ does not help. On one hand, the spectral range 
does not contain strong telluric lines, see Fig. \ref{_A6363}. However, in most cases, this line is too weak to perform the accurate 
abundance determination. Analysis of the weakest lines are impacted by the 
level of noise and accuracy of the continuum level. Uncertainties in 
the continuum at the 1\% level for the lines shown in the 
Fig. A.3 provides errors in the abundance determination of up
to 50\%. In this paper, we prefer to use the single strong Oxygen line. 
The same problem of pollution of the observed spectra by telluric 
features exists for the carbon abundance determinations. 
Carbon absorption lines that provided systematically too low or too 
high carbon abundance are excluded from the analysis.

\subsection{Solar abundance of C and O}

First of all, we reproduce the well known results of O and C abundances in the solar atmosphere.
Computations were carried out for the 1D model atmosphere with parameters 5777/4.44/0.0, where the 
oxygen and CCS line profiles were fitted using the ABEL8 program (see \citealp{pavl17}).

We fitted  profiles of the O {\sc I} and CCS lines in the spectrum of the Sun as a star \citep{kuru84} obtaining 
reasonable values for the O and C abundances. The fit to the 6300.304 \AA\ O {\sc I} line is shown in 
Fig. \ref{_AoinSun}, where we obtained a log N(O)=-3.11 with a fitted \Vm=3.48 \kmps.  
The results of the fits of the CCS lines are presented in the Table \ref{_Accs}.
In total, from the fits of all CCS lines we obtained a log N(C)=-3.59$\pm$0.01. This value agrees to 
within $\pm$0.1 dex with other previously computed values  by \cite{alex15}.

Our carbon abundance follows the form 
log N(C) = $a_{mean} \pm \sigma$,
where $a_{mean} = \sum(a_i)/N, \sigma = ((\sum (a_i-a_{mean}))/N*(N-1))^{1/2}$, $a_i$ is abundance of $i$
line from our line list, $N$ number of lines. In the case of oxygen we used ibly one line, therefore $\sigma$
cannot be determined. We estimated errors of C and O abundance determination caused by the possible inaccuracies of 
the basic parameters of the star: \Teff, \logg, \Vt for the case of the solar model \Teff/\logg = 5777/4.44, 
see Table \ref{_1}.
The majority of our stars are solar-like dwarfs, therefore their estimations can be applicable for them, as well.
As we see from the Table \ref{_1}, the variations of \Teff in 50K, \logg in 0.2, and \Vt in 0.5 provide rather
marginal changes of C and O abundances, except for the change in 0.1 dex of  log N(O) provoked by changes of \logg
by 0.2 dex.  

\begin{table}
\caption{Errors estimations for Carbon and Oxygen abundances}
\label{_1}
\begin{tabular}{cccc}
\hline\hline
Name        & \Teff $-$50K  &   \logg $-$0.2  &    \Vt$+$ 0.5  \\\hline
log N(C)     &  0.006          &   0.023           &    0.00           \\
log N(O)     &  0.020          &   0.100           &    0.00           \\
\hline

\end{tabular}
\end{table}   
 
\subsection{Dependencies log N(C) vs. log N(O) and log N(O) vs. log N(C)}

\begin{figure}
\centering
\includegraphics[width=\linewidth]{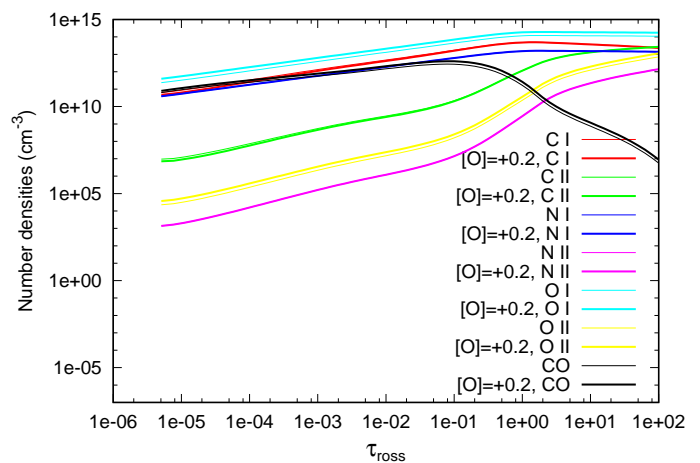}
\caption{Changes of species number densities in the solar atmosphere due to an increase of the oxygen abundance of 0.2 dex.}
\label{_co_dens}
\end{figure}

In the case of atmospheres of late-type stars, oxygen and carbon atoms form numerous molecules,
especially the CO molecule which is very stable, so abundances of O and C are bound via chemical 
equilibrium. In the case of the latest spectral classes of stars, C/O<0 and C/O>0 form
sequences of different spectral classes. For the case of solar like stars we cannot expect
large effects of the dependence of log N(C) vs. log N(O) and visa verse, yet still some effects
due to the dependence of carbon contained species, i.e. atoms C {\sc I}, C {\sc II} and molecules \ctwo vs. 
the adopted oxygen abundances, is still present, see Fig. \ref{_co_dens}.

To quantify possible effects of uncertainties in the carbon abundance determination on the results of 
the oxygen abundance measurements, and vice versa, we carried out numerical experiments for the solar model atmosphere.
Namely, we determined the oxygen abundances in the solar atmosphere for the adopted variety of 
carbon abundances, see Table \ref{_ovsc}. We found a rather marginal
response of the oxygen abundance to the variation of the carbon abundance across a broad range.

We investigated the response of the carbon abundance to the adopted oxygen abundance, see Table \ref{_ovsc}. 
Here we see notable changes of the carbon abundance at the level of 0.05 dex with changes of log N(O) of 0.5 dex, i.e.
from -2.91 to -3.41. It is worth noting that 0.05 dex is within the accuracy of our abundance determinations, estimated to be 0.1 dex.

\begin{table}
\caption{The response of the oxygen abundances in the solar atmosphere to a variety of different carbon abundances.}
\label{_ovsc}
\begin{tabular}{|c|c||c|c|}
\hline\hline
\cline{1-2}
\multicolumn{2}{|c||}{Input}& \multicolumn{2}{|c|}{Output} \\ \hline
&&&\\
log N(C) &  log N(O) &    log N(O) &  log N(C)           \\
&&&\\
\hline
&&& \\
-3.18 & -3.09        &    -2.91    &  -3.561 $\pm$ 0.013 \\
-3.28 & -3.11        &    -3.01    &  -3.579 $\pm$ 0.015 \\
-3.38 & -3.11        &    -3.11    &  -3.597 $\pm$ 0.015 \\
-3.48 & -3.11        &    -3.21    &  -3.601 $\pm$ 0.015 \\
-3.58 & -3.13        &    -3.31    &  -3.607 $\pm$ 0.015 \\
-3.68 & -3.13        &    -3.41    &  -3.611 $\pm$ 0.015 \\\hline
\end{tabular}
\end{table}

\subsection{Problems with CH lines in the blue spectral region}

Some authors use CH band heads at 4300 \AA\ to determine the carbon abundance from the fits to the spectral feature, as well as single CH lines across 
the blue spectral region, see \cite{alex15} and 
\cite{suar17}. Indeed, in the case of a well determined continuum, like in the 
solar case for instance, we obtain good agreement between results obtained with 
the fits to atomic lines + \ctwo and CH lines, which is in a good agreement 
with \cite{alex15}.

However, in our case of metal rich stars, blending becomes stronger in 
the blue part of the spectrum. Therefore, continuum determination becomes more complicated. Furthermore, in 
many cases, the S/N decreases towards the blue-end of the spectrum.
 Some of our stars have a \vsini$>$4 \kmps, increasing the uncertainty in the 
continuum level determination. Any
inaccurate continuum determination in the blue part of the HARPS spectra 
results in differences of 
abundances determined from the fits to the CH and \ctwo + C {\sc I} absorption lines. 
The same problem was noted 
by \cite{ivan17} who did not use the lines from the blue part of the CHEPS spectra. 

To reduce the uncertainties caused by problems with continuum determination in the blue part of the spectrum, we excluded CH lines from our analysis.      

\section{Results}
\label{_res}

\subsection{\vsini vs. mass}
\label{_VM}

Using empirical formulas of the dependence of mass vs. luminosity for solar-like stars \citep{eker15}, we 
computed masses for all the stars of our sample. In this case we use luminosities provided by the 
Gaia 2nd release \citep{gaia18}. As mentioned above, we used here the \vsini determined in the recent
paper by \cite{ivan17}
from the fits to Fe I line profiles.

\begin{figure}
\centering
\includegraphics[width=0.95\linewidth]{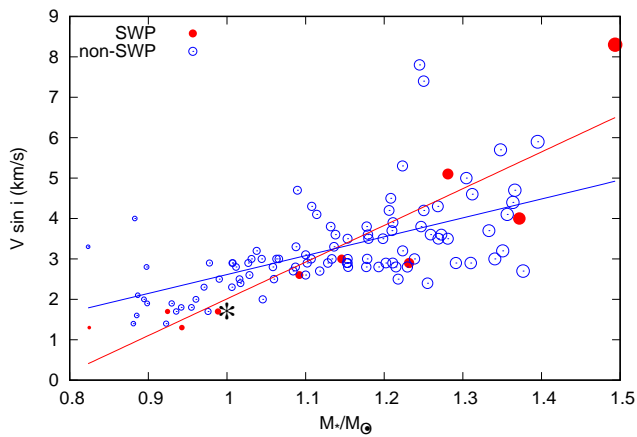}
\caption{Measured rotational velocities vs. stellar mass. Symbols sizes are a 
function of stellar mass which are given in solar masses $M_{\odot}$.}
\label{_M_vsini}
\end{figure}

We show the dependence of \vsini on stellar masses in Fig \ref{_M_vsini}. The rate of rotational momentum loss should depend on the mass of the star, 
i.e. more massive stars should remain to be fast rotators 
on longer time scales. Indeed, we confirm the dependence empirically, see 
Fig. \ref{_M_vsini}:

--  A linear approximation for the dependence of \vsini on $M_*/M_{\odot}$ is described by the formulae \vsini =$(4.68 \pm 0.65)\times(M_*/M_\odot) - 2.06 \pm 0.74$
for the non-SWP stars and
$(9.09 \pm 1.56)\times(M_*/M_\odot) - 7.08 \pm 1.793$ for the SWP stars, see blue and red lines in Fig. \ref{_M_vsini}, respectively.

-- Likely, rates of angular momentum losses are lower for more massive stars. We do not see any fast rotator at the low mass end of our sample.
The presence of more massive stars with lower \vsini can also be explained by a low sin$i$ factor, rather than a low value of $V$.

-- rates of angular momentum loss are different in SWP and non-SWP. The averaged slope of the dependence of
\vsini vs $M_*$ is steeper for the non-SWP stars. It appears that the low mass SWP lose their angular momentum 
more efficiently than low mass non-SWP.  Naturally, masses and angular momenta of proto-planetary disks are different.
However, most of our stars have evolved far enough from their time of planet formation formation for this to be an issue; also our SWP 
should have massive enough protoplanetary disks to form objects of planetary masses. For more detailed analysis we 
should carefully consider all observations made at high accuracy for all planetary systems of our SWP. This task is beyond 
the scope of this paper. Therefore, our conclusion is rather qualitative, that we observe a tendency towards lower
angular momentum for SWP in comparison to non-SWP.

It is worth noting that in this case  the  samples
of SWP and non-SWP stars are different in numbers. Our sample of SWP stars is less presentable, therefore errors are larger here, in comparison with the non-SWP stars. For more confident conclusions, the analysis of a larger number of stars is required.

\subsection{Comparison to other authors}

In general, we obtained good agreement with other authors. Due to the same 
input line lists, and well defined other parameters, i.e. effective 
temperature, gravity, abundances of other elements, our results of the 
carbon abundance determination agree with \cite{alex15} better than 0.05 dex 
for the Sun, despite some differences in the used procedure and model 
atmospheres. We obtain similar trends of [C/H] vs. [Fe/H] with \cite{niss14}, see
fig. 11 in his paper, or \cite{suar17}, see fig. 5 in their paper.

\begin{figure}
\centering
\includegraphics[width=0.95\linewidth]{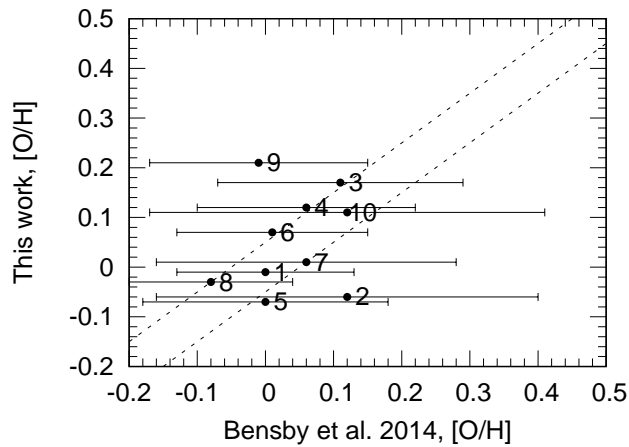}
\caption{Comparison of our oxygen abundances with \cite{bens14} for the 
common stars. Dashed lines mark the difference of abundances in $\pm$0.05 dex from the line of equal values.}
\label{_oo}
\end{figure}

In the case of the oxygen abundance results, the overall situation seems to be 
more complicated. We found 10 common stars with \cite{bens14}, they are listed 
in the Table \ref{_cb}. However, \cite{bens14} used the near-infrared 
multiplet lines of O {\sc I}
at 7771.94, 7774.17 and 7775.39 \AA\ which are known to be affected by NLTE 
effects, see \cite{kise93,aspl09}. \cite{bens14} used empirical formula to account for the NLTE corrections of O {\sc I} abundances. In this paper we used the O {\sc I} line at 6300.304 \AA\ which is less affected 
by NLTE effects. On the other hand, this line is much weaker in comparison to
the near-IR multiplet lines, and it forms a blend with a notable Ni {\sc I} line, see 
Section \ref{_oproblems}. Differences of the formally determined abundances may be on the 
order of 0.15 dex or even exceed this value, see a Table \cite{bens04}. 
Strictly speaking, the use of 
6300.304 \AA\ for O {\sc I} abundance determination is possible only in the case of known 
Ni abundance in the stellar atmosphere. Fortunately, in this paper we used 
Ni abundances in the atmospheres of our stars determined  by \cite{ivan17}.
Nevertheless, a direct comparison of our oxygen abundances for 8 out of the 10 overlapping stars 
agree within $\pm$0.05 dex, despite all differences in the input spectra and 
applied procedures for the abundance determination. The notable differences for 
two stars (NN=2 and 9, see Table \ref{_cb}) might be caused by cumulative effects
of differences in the adopted \Teff, \logg, \Vt. Besides, these stars 
are relatively fast rotators (\vsini=4.1 and 3.5 \kmps) in comparison 
to other common stars, see Table \ref{_cb}. 
The procedure of the formal approximation of the line profile by gaussian or even Voigt 
functions used by \cite{bens14} might not work correctly in some cases. 
Indeed, the rotational profile cannot be fitted by any gaussian, see \cite{gray76}.
Anyway, formally our oxygen abundances agree within the uncertainties determined 
by \cite{bens14}. De facto agreement in our abundances with \cite{bens14} 
for slowly rotating common stars is notably better, see Fig. \ref{_oo}.
 It is worth noting that the comparison with \cite{bens14} was carried out only to compare 
how the results were affected by the 
cumulative effects of differences in the procedure used, methods, and input data. 
To perform a more detailed comparison, we should follow very precisely, all the analysis steps they performed, i.e using the same line lists, damping constants, chemical 
equilibrium details, etc, as was done in \citet{ivan17}.

\begin{table*}
\caption{Comparison of the oxygen abundances in this work and \cite{bens14}.}
\label{_cb}
\begin{tabular}{cccc cccc cccc c}
\hline
\multicolumn{8}{|c|}{This work}& \multicolumn{5}{|c|}{\cite{bens14}} \\ \hline
&&&\\

NN & Name  &   \Teff&   \logg&  \Vt& \vsini&  log N(Fe)& log N(O) & \Teff&  \logg&  \Vt & log N(Fe)  & log N(O) \\ 
1 &HD 90520   &  5870  &  4.02  & 1.2 & 5.0   &-4.31 &-3.12 &  6008 & 4.16  &1.25  &-4.22  & -3.11    \\
2 & HD 150936  &  5542  &  4.13  & 1.0 & 4.1  &-4.40  &-3.17 &  5692 & 4.40  &1.12 &-4.21  & -2.99    \\
3 & HD 165204  &  5557  &  4.35  & 1.0 & 2.7  &-4.20 &-2.94 &  5637 & 4.37  &0.98  &-4.18  & -3.00    \\
4 & HD 170706  &  5698  &  4.17  & 1.0 & 3.0  &-4.24 &-2.99 &  5718 & 4.31  &0.97  &-4.24  & -3.05    \\
5 & HD 185679  &  5681  &  4.34  & 1.0 & 2.8  &-4.36 &-3.18 &  5710 & 4.47  &1.05  &-4.39  & -3.11    \\
6 & HD 186194  &  5668  &  4.09  & 1.2 & 2.9  &-4.30 &-3.04 &  5713 & 4.16  &1.06  &-4.27  & -3.10    \\
7 & HD 190125  &  5644  &  4.20  & 1.0 & 3.3  &-4.33 &-3.10 &  5682 & 4.48  &1.10  &-4.28  & -3.05    \\
8 & HD 194490  &  5854  &  4.46  & 1.0 & 3.3  &-4.41  &-3.14 &  5857 & 4.33  &1.03 &-4.39  & -3.19    \\
9 & HD 218960  &  5732  &  4.24  & 1.2 & 3.5  &-4.32 &-2.90 &  5796 & 4.09  &1.14  &-4.28  & -3.12    \\
10 & HD 220981  &  5567  &  4.34  & 1.0 & 2.4 &-4.26 &-3.00 &  5618 & 4.26  &0.96  &-4.21  & -2.99    \\
\end{tabular}
\end{table*}

\subsection{Oxygen and Carbon abundance in the CHEPS stars}

\begin{figure}
\centering
\includegraphics[width=0.95\linewidth]{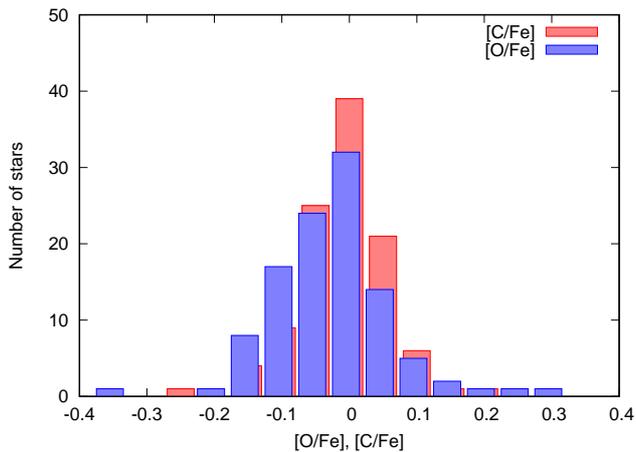}
\caption{[C/Fe] and [O/Fe] histograms for the sample of CHEPS stars, shown by different colors.}
\label{_hy}
\end{figure}

The abundances of C and O in the atmospheres of all stars of our sample are shown in Table \ref{_ACOres}. The O {\sc I} line 6300.304 \AA\ is severely blended by telluric lines in the spectra of five stars. For these stars we provide oxygen abundances scaled from the iron abundance, these cases are marked by (*) in Table \ref{_ACOres}. 
A histogram of the C and O abundance distributions is shown in Fig. \ref{_hy}.

We notice that the histogram is broader and slightly shifted for oxygen with a FWHM of 0.4 for [O/Fe] rather than 0.3 for [C/Fe]. Thus oxygen is somewhat more abundant at lower metallicities relative to carbon. It is likely that we are seeing here the results of the differing efficiency of C and O production at different times of the Galaxy’s evolution, 
if the metallicity reflects the age of the stars. 

\begin{figure}
\centering
\includegraphics[width=0.95\linewidth]{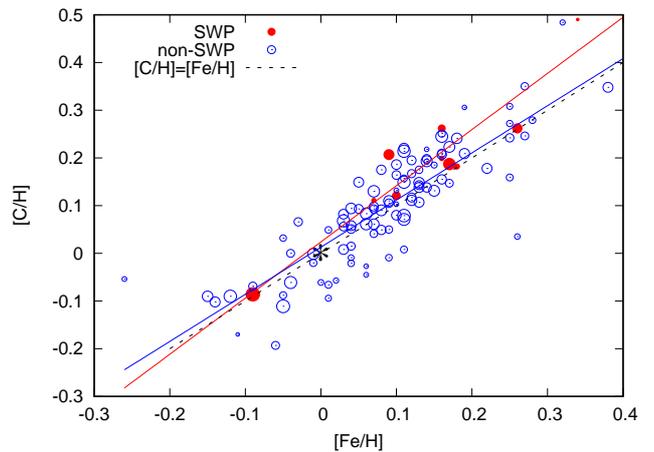}
\includegraphics[width=0.95\linewidth]{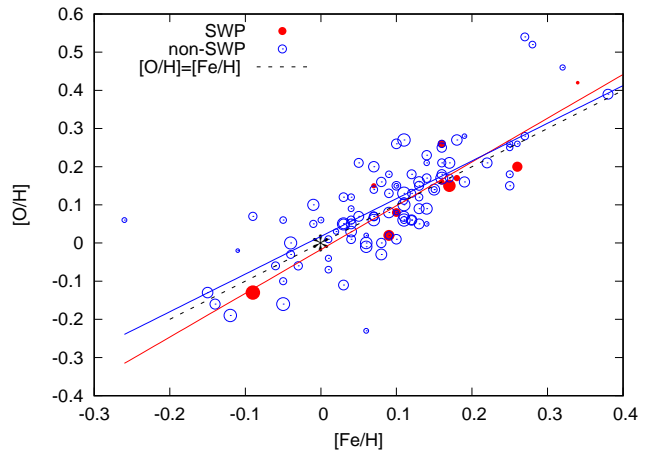}  
\caption{[C/H] and [O/H] versus [Fe/H]. The stars with planets are shown with 
red circles. Larger circles mark the stars of larger masses. The asterisk 
marks the Sun.}
\label{_CO_Fe_M}
\end{figure}

In Fig. \ref{_CO_Fe_M} we show the dependence of [O/H] and [C/H] vs. [Fe/H] in 
atmospheres of our stars. We found a notable dispersion ($\pm$0.15 dex) of the C and O abundances. Likely, 
the dispersion is intrinsic. In general, both C and O abundances follow the metallicity. Furthermore, here is no particular dependence of C and O abundance on mass.

The solar C and O abundances are always on the averaged trend. In other words, 
we do not see any peculiarity of the Sun in comparison to other stars of our sample.

 Averaged trends of oxygen and carbon abundances vs. metallicity show similar 
trends for both SWP and non-SWP. 
However, in both cases the averaged trend for non-SWP stars looks steeper. 
The difference is more pronounced in the analysis of 
[C/O], see next Sections \ref{_co}, \ref{_cov}. 

\subsection{[C/O] vs. [Fe]}
\label{_co}

\begin{figure}
\centering
\includegraphics[width=\linewidth]{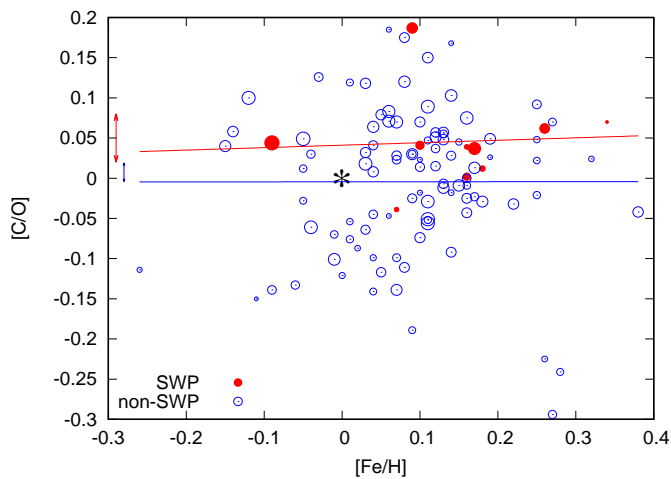}
\caption{Dependence [C/O] versus [Fe/H]. The stars with planets are shown with 
red circles. Larger circles mark the stars of larger masses. Error bars from the determinations of the mean [C/O] in the samples of SWP and non-SWP are shown 
by the red and blue arrows. Linear approximation results for the SWP and non-SWP stars shown by red ([C/O]=($(-0.576 \pm 0.273)*[Fe/H] + 0.047 \pm 0.0158 $) and blue 
([C/O] = $ (0.759 \pm 0.077)*[Fe/H] + 0.0079 \pm 0.006)$ 
lines, respectively.}
\label{_COFe_M}
\end{figure}

The differences between SWP and non-SWP are better seen in Fig. \ref{_COFe_M}.
Here we show the dependence of [C/O] vs. [Fe]. The differences between the averaged 
trends of the dependence 
are at the level of 0.05 dex, with most of our SWP having [C/O]>0, i.e. most of them are 
'carbon rich' in comparison to the solar case. 

At least three stars, i.e. HIP 29442, HIP 51987 and HIP 69724 seen in the right bottom part of the  Fig. \ref{_CO_Fe_M} manifest 
too large difference of the [C/O] with other stars. We note that only in the case of one star, HIP29442, its O {\sc I} line is severely blended by telluric line, it
 makes the oxygen abundance determination very problematic, see Fig. \ref{_A6300} and Section \ref{_oproblems}. 
 Two other stars -- HIP 51987 and HIP 69724,  show rather 'normal' carbon abundances log N(C) = -3.23, -3.20, 
  but log N(O)=-2.57 and  -2.59, respectively, see Table \ref{_ACOres}. We suggest to perform more detailed 
  study of these stars in the near future.
  
  In total, we have 10 problematic stars with the 
  poorly determined oxygen 
  abundances, see Table \ref{_ACOres}. In spectra of 
  5 stars the O {\sc I} line profile  is only partially affected by blending with telluric lines and/or 'bad pixels'. Their oxygen abundances marked by (+) were determined 'by hands', i.e. from visual comparison of computed and observed spectra.

\subsection{[C/O] vs. \vsini}
\label{_cov}

\begin{figure}
\centering
\includegraphics[width=0.95\linewidth]{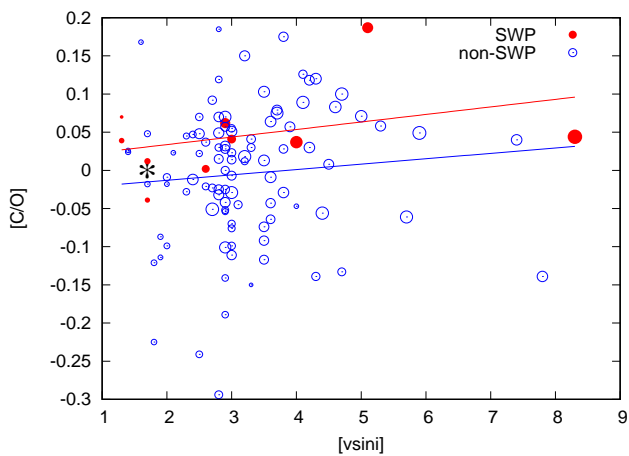}
\caption{[C/O] vs. \vsini. The Sun is marked by asterisk. More massive stars are labelled by circles of larger radius. inear approximation of results for the SWP and non-SWP stars shown by red ([C/O] = $ (0.009 \pm 0.008) \times \vsini + 0.0140 \pm 0.033) $ and blue ([C/O] = $ (  0.007 \pm 0.008)\times \vsini - 0.0271 \pm 0.028$)
lines, respectively.  }
\label{_co_vsini}
\end{figure}

Generally speaking, a fast rotation may affect the processes of matter transfer in the interiors of the star, which could give rise to a dependence of [C/O] vs. \vsini. 
However, our results do not confirm this suggestion: in Fig. \ref{_co_vsini} 
we see only a slight increase of [C/O] toward larger \vsini. However, our stars are comparatively slow rotators by selection, therefore this hypothesis should be tested on a larger sample of stars across a wider range of \vsini.

\onecolumn
\begin{longtable}{lccc cccc ccl}
\caption{Carbon and Oxygen in the CHEPS stars. Cases of the problematic determination of oxygen abundance  due to the strong blending 6300.304 \AA~ O {\sc I} line by telluric lines are marked by (*). The oxygen abundances determined by visual comparison of computed and observed spectra are marked by (+).
\label{_ACOres}} \\
\hline\hline \\
\\
SWP     & \Teff   &   \logg &    \Vt  &  \vsini    &  M/M$_{\odot}$      &  L/L$_{\odot}$       &      Age(Gyr)      &   [Fe/H] &       [C/H]           &   [O/H]     \\
\\ \hline                                                                                            
\endfirsthead
\hline\hline \\
\\
SWP     & \Teff   &   \logg &    \Vt  &  \vsini    &  M/M$_{\odot}$      &  L/L$_{\odot}$       &      Age(Gyr)      &   [Fe/H] &       [C/H]           &   [O/H]     \\
\\ \hline                                                                                            
\endhead
\hline
\endfoot
HD9174      &  5577   &   4.05  &    1.2  &    2.9       &   1.231     & 2.581    &       5.8     &    0.26  &      0.262$\pm$0.021    &    0.20       \\
HD48265     &  5651   &   3.92  &    1.4  &    4.0       &   1.371     & 4.350    &       4.0     &    0.17  &      0.187$\pm$0.030    &    0.15       \\
HD77338     &  5315   &   4.42  &    1.2  &    1.3       &   0.942     & 0.708    &       9.5     &    0.16  &      0.199$\pm$0.030    &    0.16       \\
HD128356    &  4875   &   4.58  &    0.8  &    1.3       &   0.824     & 0.371    &      15.5     &    0.34  &      0.490$\pm$0.038    &    0.42       \\
HD143361    &  5505   &   4.42  &    1.0  &    1.7       &   0.988     & 0.891    &       8.5     &    0.18  &      0.182$\pm$0.016    &    0.17       \\
HD147873    &  5972   &   3.90  &    1.4  &    8.3       &   1.493     & 6.568    &       2.6     &   -0.09  &     -0.086$\pm$0.027    &   -0.13       \\
HD152079   &  5726   &   4.35  &    1.2  &    2.6       &   1.092     & 1.443    &       6.2     &    0.16  &      0.262$\pm$0.014    &    0.26(+)       \\
HD154672    &  5655   &   4.16  &    1.2  &    3.0       &   1.145     & 1.820    &       5.0     &    0.10  &      0.121$\pm$0.014    &    0.08       \\
HD165155    &  5426   &   4.57  &    1.0  &    1.7       &   0.924     & 0.644    &       8.5     &    0.07  &      0.111$\pm$0.011    &    0.15       \\
HD224538    &  6097   &   4.29  &    1.4  &    5.1       &   1.281     & 3.124    &       4.0     &    0.09  &      0.207$\pm$0.023    &    0.02       \\
&&&& &&&& &&\\ \hline
&&&& &&&& &&\\
 \hline
&&&& &&&& &&\\
non-SWP   & \Teff   &   \logg &   \Vt  &  \vsini &   M/M$_{\odot}$   &    L/L$_{\odot}$     &    Age(Gyr)     &    [Fe/H] &   [C/H]           &   [O/H]     \\
&&&& &&&& &&\\ \hline                                                                                  
&&&& &&&& &&\\                                                                                         
HD6790     &  6012   &   4.40  &    0.8  &    4.7   &  1.089    &  1.427   &     3.5    &   -0.06  &    -0.193$\pm$0.017    &   -0.06(*)     \\
HD7950      &  5426   &   3.94  &    1.2  &    2.7   & 1.376    &  4.426   &     4.0    &     0.11  &     0.079$\pm$0.017    &    0.13       \\
HD8389      &  5243   &   4.52  &    1.2  &    1.4   & 0.922    &  0.638   &    11.5    &     0.32  &     0.484$\pm$0.029    &    0.46       \\
HD8446      &  5819   &   4.14  &    1.2  &    3.9   & 1.211    &  2.380   &     5.6    &     0.13  &     0.107$\pm$0.018    &    0.05       \\
HD10188     &  5714   &   4.16  &    1.2  &    3.8   & 1.246    &  2.740   &     5.4    &     0.18  &     0.241$\pm$0.030    &    0.27       \\
HD10278     &  5712   &   4.62  &    1.0  &    2.9   & 0.977    &  0.845   &     6.0    &     0.01  &    -0.094$\pm$0.012    &   -0.04       \\
HD13147     &  5502   &   3.94  &    1.0  &    2.9   & 1.202    &  2.295   &     4.0    &     0.03  &     0.082$\pm$0.023    &    0.05       \\
HD13350     &  5515   &   4.22  &    1.2  &    2.7   & 1.118    &  1.616   &     6.8    &     0.25  &     0.242$\pm$0.029    &    0.15       \\
HD15507     &  5766   &   4.62  &    1.0  &    2.9   & 1.006    &  0.973   &     7.0    &     0.09  &    -0.009$\pm$0.014    &    0.18       \\
HD18708     &  5838   &   4.36  &    1.2  &    3.6   & 1.138    &  1.761   &     4.2    &     0.03  &     0.056$\pm$0.017    &    0.12       \\
HD18754     &  5531   &   3.84  &    1.4  &    3.2   & 1.350    &  4.036   &     3.8    &     0.03  &     0.068$\pm$0.020    &    0.05       \\
HD19493     &  5743   &   4.16  &    1.2  &    3.0   & 1.238    &  2.653   &     5.1    &     0.12  &     0.111$\pm$0.019    &    0.06       \\
HD19773     &  6156   &   4.13  &    1.2  &    4.2   & 1.206    &  2.334   &     3.9    &     0.03  &     0.008$\pm$0.010    &   -0.11       \\
HD23398    &  5592   &   4.10  &    1.2  &    2.9   & 1.210    &  2.378   &     6.4    &     0.38  &     0.348$\pm$0.020    &    0.38(*)       \\
HD26071     &  5549   &   4.16  &    1.2  &    2.8   & 1.154    &  1.884   &     5.2    &     0.12  &     0.195$\pm$0.024    &    0.18       \\
HD29231     &  5400   &   4.43  &    1.2  &    1.9   & 0.929    &  0.662   &     7.0    &     0.02  &    -0.057$\pm$0.017    &    0.03       \\
HD38459     &  5233   &   4.43  &    1.2  &    4.0   & 0.882    &  0.515   &     9.0    &     0.06  &    -0.027$\pm$0.026    &    0.02       \\
HD38467     &  5721   &   4.18  &    1.2  &    2.8   & 1.192    &  2.211   &     5.6    &     0.10  &     0.080$\pm$0.020    &    0.01       \\
HD40293     &  5549   &   4.51  &    1.0  &    1.8   & 0.954    &  0.753   &     6.0    &     0.00  &    -0.061$\pm$0.016    &    0.06       \\
HD42538     &  5939   &   3.98  &    1.4  &    5.7   & 1.347    &  3.997   &     3.6    &    -0.04  &    -0.061$\pm$0.018    &    0.00       \\
HD42719     &  5809   &   4.08  &    1.4  &    4.4   & 1.363    &  4.225   &     3.8    &     0.11  &     0.214$\pm$0.027    &    0.27       \\
HD42936     &  5126   &   4.44  &    0.8  &    1.4   & 0.881    &  0.510   &    12.0    &     0.19  &     0.306$\pm$0.026    &    0.28       \\
HD45133     &  5601   &   4.31  &    1.2  &    2.9   & 1.102    &  1.507   &     6.1    &     0.16  &     0.210$\pm$0.023    &    0.21       \\
HD49866     &  5712   &   3.71  &    1.4  &    4.7   & 1.366    &  4.264   &     3.3    &    -0.12  &    -0.090$\pm$0.027    &   -0.19       \\
HD50652    &  5641   &   4.21  &    1.2  &    2.9   & 1.152    &  1.871   &     5.2    &     0.12  &     0.117$\pm$0.017    &    0.06(+)       \\
HD55524     &  5700   &   4.22  &    1.2  &    3.5   & 1.180    &  2.099   &     6.0    &     0.14  &     0.138$\pm$0.023    &    0.23       \\
HD56259     &  5489   &   3.94  &    1.2  &    3.0   & 1.340    &  3.894   &     4.4    &     0.11  &     0.071$\pm$0.015    &    0.10       \\
HD56413     &  5648   &   4.41  &    1.2  &    2.9   & 1.027    &  1.072   &     6.6    &     0.11  &     0.008$\pm$0.015    &    0.06       \\
HD56957     &  5674   &   4.09  &    1.4  &    3.5   & 1.280    &  3.120   &     4.8    &     0.14  &     0.193$\pm$0.031    &    0.09       \\
HD61475     &  5250   &   4.47  &    1.2  &    2.0   & 0.894    &  0.549   &     9.0    &     0.10  &     0.132$\pm$0.018    &    0.15       \\
HD66653     &  5771   &   4.42  &    1.2  &    3.2   & 1.037    &  1.127   &     4.2    &    -0.05  &    -0.088$\pm$0.020    &   -0.10       \\
HD69721     &  5296   &   4.47  &    1.0  &    1.7   & 0.935    &  0.682   &     9.0    &     0.14  &     0.192$\pm$0.029    &    0.21       \\
HD76849    &  5223   &   5.00  &    1.0  &    1.9   & 0.898    &  0.561   &     6.4    &    -0.26  &    -0.054$\pm$0.039    &    0.06(+)       \\
HD78130     &  5744   &   4.43  &    1.0  &    2.9   & 1.007    &  0.976   &     6.1    &     0.04  &    -0.021$\pm$0.012    &    0.12       \\
HD78286     &  5794   &   4.40  &    1.2  &    2.9   & 1.128    &  1.689   &     5.2    &     0.09  &     0.105$\pm$0.014    &    0.13       \\
HD86006     &  5668   &   3.97  &    1.2  &    3.6   & 1.272    &  3.026   &     5.0    &     0.15  &     0.131$\pm$0.019    &    0.14       \\
HD90028     &  5740   &   4.06  &    1.2  &    4.2   & 1.250    &  2.776   &     5.0    &     0.09  &     0.110$\pm$0.027    &    0.08       \\
HD90520     &  5870   &   4.08  &    1.2  &    5.0   & 1.304    &  3.409   &     4.3    &     0.06  &     0.061$\pm$0.024    &   -0.01       \\
HD91682     &  5614   &   4.13  &    1.2  &    3.0   & 1.178    &  2.084   &     5.6    &     0.08  &     0.049$\pm$0.020    &    0.16       \\
HD93849     &  6153   &   4.21  &    1.2  &    4.3   & 1.268    &  2.974   &     3.5    &     0.08  &     0.090$\pm$0.015    &   -0.03       \\
HD95136     &  5744   &   4.41  &    1.2  &    3.1   & 1.100    &  1.494   &     5.9    &     0.04  &     0.015$\pm$0.023    &    0.06       \\
HD96494     &  5356   &   4.53  &    1.0  &    2.8   & 0.897    &  0.559   &     9.0    &     0.06  &    -0.045$\pm$0.022    &   -0.23       \\
HD101197    &  5756   &   4.23  &    1.0  &    4.5   & 1.208    &  2.354   &     4.2    &     0.04  &     0.058$\pm$0.039    &    0.05       \\
HD101348    &  5620   &   3.95  &    1.4  &    4.1   & 1.356    &  4.118   &     4.0    &     0.11  &     0.149$\pm$0.025    &    0.06       \\
HD102196    &  6012   &   3.90  &    1.4  &    5.9   & 1.395    &  4.719   &     3.0    &    -0.05  &    -0.111$\pm$0.013    &   -0.16       \\
HD102361    &  5978   &   4.12  &    1.4  &    7.4   & 1.250    &  2.775   &     2.0    &    -0.15  &    -0.090$\pm$0.051    &   -0.13       \\
HD105750    &  5672   &   4.24  &    0.8  &    4.6   & 1.311    &  3.506   &     4.3    &     0.06  &     0.083$\pm$0.025    &    0.00       \\
HD106937    &  5455   &   4.05  &    1.2  &    2.5   & 1.217    &  2.443   &     5.5    &     0.13  &     0.138$\pm$0.018    &    0.09       \\
HD107181    &  5581   &   4.17  &    1.2  &    2.8   & 1.214    &  2.416   &     6.0    &     0.22  &     0.178$\pm$0.022    &    0.21       \\
HD108953    &  5514   &   4.43  &    1.0  &    2.5   & 0.990    &  0.898   &     9.3    &     0.25  &     0.272$\pm$0.025    &    0.25       \\
HD126535    &  5284   &   4.65  &    1.0  &    2.1   & 0.886    &  0.526   &     9.0    &     0.10  &     0.103$\pm$0.020    &    0.08       \\
HD127423    &  6020   &   4.26  &    1.0  &    4.3   & 1.107    &  1.546   &      3.1   &    -0.09  &    -0.069$\pm$0.018    &    0.07       \\
HD143120   &  5576   &   3.95  &    1.2  &    3.7   & 1.333    &  3.788   &     4.4    &     0.16  &     0.245$\pm$0.022    &    0.17(+)       \\
HD144550    &  5652   &   4.19  &    1.2  &    3.8   & 1.177    &  2.078   &     5.6    &     0.08  &     0.175$\pm$0.030    &    0.00       \\
HD144848    &  5777   &   4.26  &    1.2  &    3.5   & 1.198    &  2.261   &     5.6    &     0.10  &     0.186$\pm$0.016    &    0.26       \\
HD144899    &  5833   &   4.13  &    1.2  &    3.5   & 1.268    &  2.982   &     5.0    &     0.17  &     0.223$\pm$0.019    &    0.21       \\
HD149189    &  5771   &   4.08  &    1.4  &    3.6   & 1.258    &  2.871   &     4.6    &     0.04  &     0.094$\pm$0.027    &    0.03       \\
HD149782    &  5554   &   4.34  &    1.2  &    2.3   & 1.006    &  0.971   &     4.5    &    -0.05  &     0.032$\pm$0.023    &    0.06       \\
HD150936    &  5542   &   4.12  &    1.0  &    4.1   & 1.114    &  1.590   &     3.6    &    -0.03  &     0.066$\pm$0.036    &   -0.06       \\
HD154221   &  5797   &   4.47  &    1.0  &    3.0   & 1.031    &  1.093   &     5.3    &     0.01  &    -0.066$\pm$0.012    &    0.01(*)       \\
HD158469    &  6105   &   4.19  &    1.2  &    5.3   & 1.223    &  2.498   &     2.0    &    -0.14  &    -0.102$\pm$0.019    &   -0.16       \\
HD165204   &  5557   &   4.33  &    1.0  &    2.7   & 1.084    &  1.393   &     6.5    &     0.17  &     0.147$\pm$0.020    &    0.17(*)       \\
HD170706    &  5698   &   4.40  &    1.0  &    3.0   & 1.063    &  1.266   &     6.2    &     0.13  &     0.175$\pm$0.020    &    0.12       \\
HD178340    &  5538   &   4.38  &    1.2  &    2.0   & 1.045    &  1.168   &     7.0    &     0.16  &     0.251$\pm$0.019    &    0.26       \\
HD178787    &  5216   &   4.44  &    1.0  &    1.6   & 0.885    &  0.522   &    11.0    &     0.14  &     0.218$\pm$0.023    &    0.05       \\
HD185679    &  5681   &   4.43  &    1.0  &    2.8   & 1.011    &  0.996   &     5.5    &     0.01  &     0.049$\pm$0.015    &   -0.07       \\
HD186194   &  5668   &   4.30  &    1.2  &    2.9   & 1.153    &  1.881   &     4.8    &     0.07  &     0.098$\pm$0.018    &    0.07(*)       \\
HD186265    &  5562   &   4.39  &    1.2  &    2.5   & 1.059    &  1.247   &     8.0    &     0.27  &     0.350$\pm$0.026    &    0.28       \\
HD189627    &  6210   &   4.40  &    1.4  &    7.8   & 1.244    &  2.719   &     4.0    &    0.07  &     0.061$\pm$0.018    &    0.20       \\
HD190125    &  5644   &   4.53  &    1.0  &    3.3   & 1.136    &  1.747   &     5.0    &     0.04  &     0.051$\pm$0.020    &    0.01       \\
HD191122    &  5851   &   4.34  &    1.2  &    3.0   & 1.133    &  1.727   &     5.1    &     0.10  &     0.164$\pm$0.017    &    0.15       \\
HD191760   &  5816   &   4.10  &    1.4  &    2.9   & 1.310    &  3.481   &     4.2    &     0.07  &     0.130$\pm$0.024    &    0.06(+)       \\
HD193690    &  5558   &   4.48  &    1.0  &    2.3   & 0.970    &  0.816   &     8.6    &     0.15  &     0.185$\pm$0.015    &    0.14       \\
HD193995    &  5661   &   4.09  &    1.2  &    3.2   & 1.223    &  2.499   &     5.4    &     0.11  &     0.220$\pm$0.019    &    0.07       \\
HD194490    &  5854   &   4.44  &    1.0  &    3.3   & 1.087    &  1.415   &     3.8    &    -0.04  &     0.000$\pm$0.015    &   -0.03       \\
HD200869    &  5401   &   4.37  &    1.0  &    1.7   & 0.975    &  0.837   &     9.9    &     0.25  &     0.308$\pm$0.023    &    0.26       \\
HD201757    &  5597   &   4.23  &    1.2  &    3.7   & 1.209    &  2.367   &     5.0    &     0.05  &     0.149$\pm$0.039    &    0.07       \\
HD206683    &  5909   &   4.37  &    1.2  &    3.8   & 1.131    &  1.716   &     5.5    &     0.14  &     0.198$\pm$0.015    &    0.17       \\
HD206837    &  5616   &   4.07  &    1.2  &    2.9   & 1.291    &  3.245   &     4.2    &    -0.01  &    -0.001$\pm$0.019    &    0.10       \\
HD218960    &  5732   &   4.27  &    1.2  &    3.5   & 1.153    &  1.879   &     4.4    &     0.05  &     0.093$\pm$0.022    &    0.21       \\
HD219011    &  5642   &   4.21  &    1.2  &    3.0   & 1.153    &  1.880   &     5.4    &     0.13  &     0.143$\pm$0.024    &    0.15       \\
HD219556    &  5485   &   4.44  &    1.0  &    2.0   & 0.960    &  0.775   &     7.2    &     0.04  &    -0.009$\pm$0.019    &    0.09       \\
HD220981    &  5567   &   4.33  &    1.0  &    2.4   & 1.017    &  1.023   &     7.0    &     0.11  &     0.157$\pm$0.027    &    0.11       \\
HD221575    &  5037   &   4.49  &    1.4  &    3.3   & 0.823    &  0.368   &     6.0    &    -0.11  &    -0.170$\pm$0.049    &   -0.02       \\
HD221954    &  5602   &   4.10  &    1.2  &    2.8   & 1.229    &  2.561   &     5.6    &     0.19  &     0.209$\pm$0.022    &    0.16       \\
HD222910    &  5480   &   4.05  &    1.2  &    2.4   & 1.255    &  2.828   &     5.1    &     0.13  &     0.148$\pm$0.021    &    0.16       \\
HIP19807    &  5892   &   4.54  &    1.0  &    3.0   & 1.065    &  1.283   &     5.5    &     0.07  &     0.041$\pm$0.011    &    0.14       \\
HIP28641    &  5747   &   4.45  &    1.0  &    3.0   & 1.044    &  1.161   &     4.8    &    -0.01  &    -0.020$\pm$0.015    &    0.05       \\
HIP29442   &  5322   &   4.43  &    0.8  &    1.8   & 0.941    &  0.705   &    11.0    &     0.26  &     0.035$\pm$0.025    &    0.26(*)       \\
HIP31831    &  5845   &   4.29  &    1.2  &    3.6   & 1.179    &  2.092   &     5.8    &     0.16  &     0.207$\pm$0.022    &    0.25       \\
HIP43267    &  5642   &   4.29  &    1.2  &    2.6   & 1.100    &  1.494   &     5.8    &     0.12  &     0.167$\pm$0.027    &    0.13       \\
HIP51987    &  6158   &   5.10  &    1.0  &    2.8   & 1.087    &  1.411   &     7.2    &     0.27  &     0.246$\pm$0.027    &    0.54       \\
HIP53084    &  5527   &   4.32  &    1.0  &    2.6   & 1.028    &  1.079   &     8.4    &     0.25  &     0.159$\pm$0.033    &    0.18       \\
HIP57331    &  5531   &   4.21  &    1.2  &    2.8   & 1.058    &  1.240   &     6.1    &     0.09  &     0.050$\pm$0.028    &    0.02       \\
HIP66990    &  5595   &   4.15  &    1.2  &    2.8   & 1.177    &  2.075   &     6.1    &     0.16  &     0.155$\pm$0.026    &    0.18       \\
HIP69724    &  5793   &   4.79  &    1.0  &    2.5   & 1.016    &  1.017   &     9.0    &     0.28  &     0.279$\pm$0.017    &    0.52       \\
HIP111286  &  5690   &   4.19  &    1.2  &    3.0   & 1.107    &  1.542   &     5.1    &     0.07  &     0.093$\pm$0.025    &    0.07(*)       \\
&&&& &&&& \\ \hline                                                           
\end{longtable}
\twocolumn

On the other hand, we see here the results discussed in the former section, 
see \ref{_VM}. Namely, almost all our massive stars 
occupy the range of \vsini$>$3 \kmps, regardless of whether they are SWP or non-SWP.

\subsection{Lithium vs. Oxygen}

 Lithium is an element of special interest in many aspects of modern astrophysics, see \cite{pavl18} and
references therein. Li is a fragile element,
theoretical models predict lithium depletion that can be compared to the lithium in the Sun and stars of different ages and 
metallicities. \cite{piau02} distinguished two phases in lithium depletion: (1) a rapid nuclear destruction in the T Tauri phase before 20 Myr in stars of 0.8 and 1.4 M$\odot$ masses, which is dependent on the extent and temperature of the convective zone, and (2) a second phase where the destruction is slow and moderate and which is largely dependent on the (magneto)hydrodynamic instability located at the base of the convective zone. 
\cite{piau02} outlined the importance of the O/Fe ratio, due to the role of oxygen as known source of opacity. 

\begin{figure}
\centering
\includegraphics[width=0.95\linewidth]{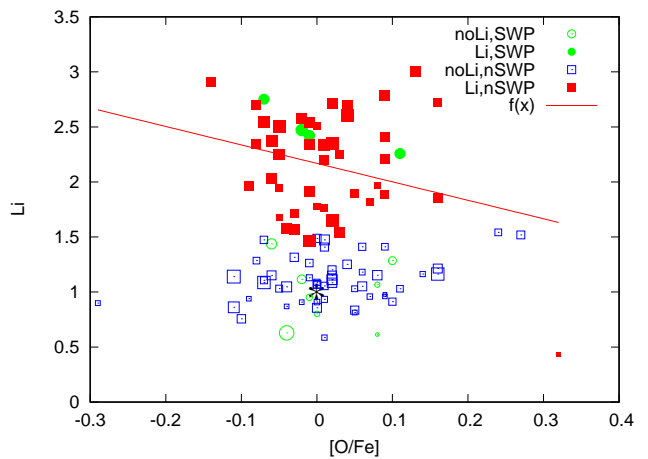}
\caption{[Li] versus [O] for stars with a detectable lithium 6708\AA\ line, and the upper limit of lithium
abundances in stars of our sample. The Sun is marked by an asterisk. More massive stars 
are labeled by figures of larger radius.}
\label{_Li_O_M2}
\end{figure}

To investigate the problem of the possible  
dependence of the lithium abundance vs. the oxygen abundance, we used the results of the 
Li determinations by \cite{pavl18}, where the dependence of N(Li) vs. N(O) is shown in Fig. \ref{_Li_O_M2} 
for all stars of our sample, i.e.  with detectable Li, and the upper limits of lithium abundances.
It is clear that our sample is not complete enough to provide a firm conclusion, 
but there does appear the notable connection between the lithium and oxygen abundances, shown by the red line in Fig. \ref{_Li_O_M2}, i.e log N(Li) = 2.16 $\pm$ 0.07 -
 (1.67$\pm$0.9)*[O/Fe]. The slope of the dependence is determined with a large error bar, but it is negative here,
 in accordance with the prediction of \cite{piau02}. 
Interestingly, all four of our SWP are found 
 to have lithium abundances above the mean linear fit to the data.

\subsection{O and C in stars of different age}

 Using the luminosities of our stars from the Gaia DR2 release, along with 
the recent evolutionary tracks from the MIST group,
decribed in detail by \cite{dott16, choi16, paxt11, paxt13, paxt15},
we determined the ages of our stars.
Generally speaking, measuring the ages of low mass stars to high precision is difficult, particularly 
old Sun-like stars on the MS, where evolutionary model convergence can 
give rise to large uncertainties. Low mass stars, i.e stars of solar 
masses or smaller, evolve very slowly, such that 
their luminosities increase slowly with time, as seen in Fig. \ref{_Aagesmu}.
Furthermore, evolution of stars notably depends on their metallicities, whereby 
metal deficient stars are generally of a higher luminosity, and therefore they
evolve in shorter time scales. Uncertainties in the  metallicity of $\pm$ 0.2  dex can give rise to an 
uncertainty of $\pm$ 2 Myr in age, for a star of solar mass.
We computed evolutionary tracks of our stars for their masses and metallicities
given in the Table \ref{_ACOres} using the Web Interpolator
(http://waps.cfa.harvard.edu/MIST/)
of the MIST database. The comparison of the observed luminosities and theoretical 
MIST luminosities are available on 
ftp://ftp.mao.kiev.ua/pub/yp/2018/CandO/tagesL.pdf.
Ages of our stars are given in the Table \ref{_ACOres}.
We note that for some stars, we cannot find proper solutions, meaning their calculated ages exceed a Hubble time, i.e $>$14 Gyr.
We highlight them in our plots to underline the existing problems in age
determinations using even the latest and most up-to-date evolutionary tracks. 
In any case, the majority of these 'strange' stars are of masses lower than 1M$_{\odot}$.
Detailed analysis of these results is beyond the scope of this paper, but is an avenue that should be explored further in the future.

Nevertheless, we show the dependence [C/Fe] and [O/Fe] vs. age in Figs. \ref{_AaC} and
\ref{_AaO}, respectively. In both cases we see rather positive slopes of the
dependencies. It can be interpreted as evidence that at older times, 
production of C and O prevails over Fe yield.

\section{Discussion}
\label{_dis}

We carried out our analysis in the framework of classical approach. Since the stars of our sample are dwarfs, we employed the simplifications of using LTE, 1D, and no sinks and energy sources in the atmosphere, therefore 
mixing-length theory of convection
 is still valid for our model computations. Indeed, for the case of the solar atmosphere we obtain a good agreement with other authors for the investigated abundances, \Vt and \vsini.

In this paper we adopt the \cite{ande89} abundance scale. This was done only to fix 
the 'zero point' in our computations. Generally speaking, we performed our abundance and other 
parameter analyses relative to the Sun.
In other words, our procedure was adopted for the Sun, and then later we performed the same 
analysis on other stars. Our analysis shows that the Sun can be considered a normal star 
in this respect, since its parameters correspond well 
to the average parameters of the stars in our sample.

In the framework of the project, we determine masses of all stars of our 
sample using luminosities provided by the second data release of Gaia, 
allowing us to add one more dimension to our analysis. At least 10 of 
our stars are SWP. The comparison of SWP and non-SWP provides 
another interesting aspect of our research. It is worth noting that the analysis 
of SWP and non-SWP was carried out in the framework of the same approach. 
In other words, all abundances and other parameters were determined using fits to 
the observed spectra observed by one instrument, and the the same procedure 
of analysis was employed.

 Despite the problems of age determination of G-dwarfs we computed the ages of the stars
of our sample using the comparison of Gaia DR2 luminosities with the MIST evolutionary tracks.
The accuracy of age determination of lower mass stars seems to be controversial in many cases
though both dependencies [C/Fe] and [O/Fe] vs. age are positive. Likely, the production 
of C and O was more effective in comparison with Fe in former epochs.

Usually, in abundance analysis we use many lines to reduce possible formal errors. 
In the case of the oxygen abundance determination we used only one line at 6300.304 \AA. 
However, it has been studied in detail by many authors, \citep[see][]{alle01}. 
The line blends with the Ni {\sc I} line at 6300.336 \AA, which practically means the 
use of the 6300.304 \AA\ line is restricted to cases where the Ni abundance is 
known. Fortunately, the Ni abundance in the stars of our sample was 
previously determined \citep{ivan17}. Furthermore, we adjusted the parameters of the Ni 
line to obtain the known solar abundance \citep{ande89}. This allows us to 
reproduce the O {\sc I} and Ni {\sc I} line blend in the solar spectrum.
We do not use the 6363 \AA\ line due to its weakness. Unfortunately, even 
small uncertainties across weak spectral lines
reduce the accuracy of abundance determination from the fit to this line.
Other important input parameters, i.e. effective temperatures, gravities, micro- 
and microturbulent velocities were taken from \cite{ivan17}.

Carbon abundances were determined using the list of C {\sc I} and \ctwo lines from the 
paper of \cite{alex15}. Our results for the Sun agree well with these authors 
within $\pm$0.02 dex, despite the differences in model atmospheres, 
procedures, etc. We do not use lines from the CH molecule that appear in the 
blue spectral region due to strong blending there which affects the 
continuum determination.

The \vsini for all stars of our sample was determined by \cite{ivan17}, and we 
found that most of our fast rotators are more massive stars.
 
From our analysis we do not find evidence that the [C/O] ratio depends 
on \vsini or mass of the star.
We do not find any correlation between [C/O] and the lithium abundance.
On the other hand, we see a large dispersion of C and O abundances for the stars of different masses and metallicities.
Even if we could explain some of the observed dispersion of oxygen 
from the use of only a single line, the problem 
of the abundance dispersion still remains for carbon, which was determined by the 
analysis of a large enough number of atomic and molecular lines.

Despite the notable dispersion of C and O abundances, our histogram analysis 
showed some differences in the 
distribution of these species. 

In general, carbon and oxygen abundances follow metallicity. However, the comparison between their overall distributions show that more oxygen was formed at lower metallicities, whereas carbon formation was more effective at higher metallicities and hence – later times, if metallicity reflects the time of formation of the star. Thus carbon production has been more effective at later epochs in the evolution of the Milky Way.
In this way we see the slightly different abundance histories of oxygen and carbon in our Galaxy.

We note that FHWMs of [C/Fe[ and [O/Fe] distributions in atmospheres of stars of our sample are subtly different, with oxygen production being somewhat favoured at lower metallicities relatively to Carbon. Our targets are mainly G-dwarfs, which cannot easily change their surface abundances of oxygen and carbon on short time scales.  This presuably arises from the differences in the processes yielding C and O at different metallicities during the evolution of our Galaxy for which is some evidence from other work (e.g. \cite{chia03}).

Finally,  we confirm that SWP are more carbon rich in comparison with non-SWP,
in agreement with e.g., \cite{suar17}.
It is worth noting that this result was obtained from 
analysis of only 10 SWP and 97 non-SWP stars. On the other hand, our analysis was done in 
the framework of a homogeneous set of the observed data, all using
the same procedure. Therefore, this abundance difference between SWP and non-SWP is real. At least, the shift between averaged [C/O] 
for the samples of SWP and non-SWP exceeds the formal error bars determined for these samples.

We carry out numerical Monte-Carlo simulations of the [C/Fe] distributions 
to estimate the reality of the computed differences for SWP and non-SWP, see Fig. \ref{_AMC}.
Random numbers were generated by the RANDOM\_NUMBERS procedure of Fortran 95, and an extended set of Monte-Carlo simulations (N=10$^8$) was carried out, drawing from normal distributions of the [C/Fe] abundance ratio in both SWP and non-SWP samples. 
A normal distribution was used here based on the results for [C/Fe] we found in Fig. \ref{_hy}. 
We found a probability of the null result of only 3.9 \%. In other words, there is a 96.1 \% probability that the populations are drawn from different parent distributions, a value approaching 3$\sigma$.

\section{Acknowledgements}

 We  acknowledge support by Fondecyt grant 1161218 and partial support by CATA-Basal (PB06, CONICYT) and support from the UK STFC via grants ST/M001008/1. YP's work has been supported in part by grant from the Polish National Science Center 2016/21/B/ST9/01626.
 Special thanks to developers of the MIST evolutionary tracks database.
We thank the anonymous Referee for their thorough review and highly appreciate the comments and suggestions, which significantly contributed to improving the quality of the publication.

\bibliographystyle{aa}
\bibliography{mnemonic,bib_C}

\newpage
\clearpage
\begin{appendix}
\section{}

\begin{table}
\caption{Spectroscopic parameters of the absorption lines which form the 6300.3 \AA\ feature.}
\label{_AOIl}
\begin{tabular}{cccc}

Atom     & $\lambda$& $gf$      & E''   \\        
    Ce {\sc I} & 6300.204 & 6.124E-02 & 0.159 \\
   Cr {\sc I}  & 6300.256 & 3.793E-05 & 4.613 \\
   Ce {\sc II} & 6300.265 & 3.467E-03 & 3.822 \\ 
     O {\sc I} & 6300.304 & 1.517E-10 & 0.000 \\
    Ni {\sc I} & 6300.336 & 2.500E-03 & 4.266 \\     
    Tm {\sc I} & 6300.355 & 2.042E-03 & 2.692 \\
    Fe {\sc I} & 6300.415 & 3.864E-04 & 5.033 \\
\end{tabular}
\end{table}

\begin{figure}
\centering
\includegraphics[width=0.95\linewidth]{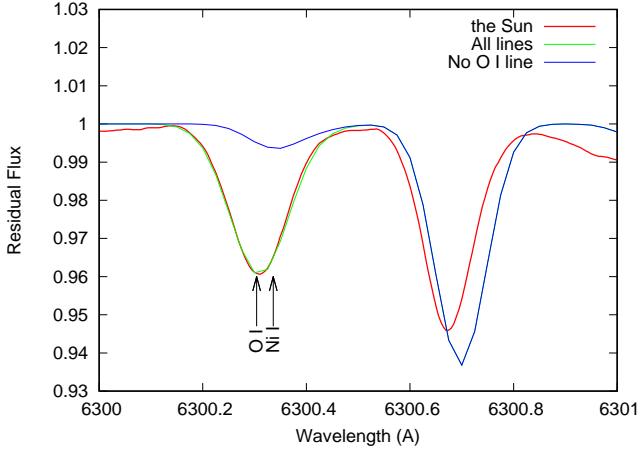}
\caption{Contribution of O {\sc I}, Ni {\sc I} and other lines into the formation of the 6300.304 \AA\ line in the solar spectrum.}
\label{_AoinSun}
\end{figure}

\begin{figure}
\centering
\includegraphics[width=0.95\linewidth]{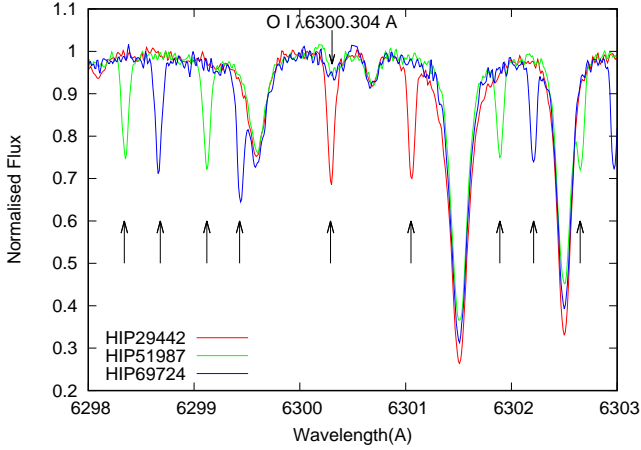}
\caption{Spectral range around the O {\sc I} $\lambda$ 6300.304 \AA\ line observed for three stars of our CHEPS sample. Telluric O$_2$ lines are marked by arrows.}
\label{_A6300}
\end{figure}

\begin{figure}
\centering
\includegraphics[width=0.95\linewidth]{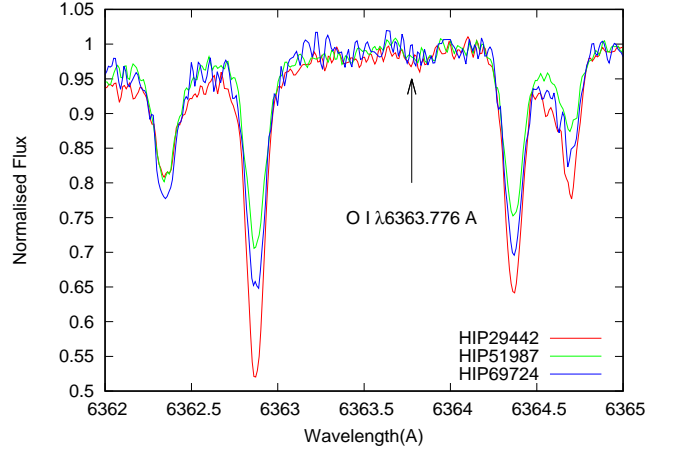}
\caption{Spectral range around the O {\sc I} $\lambda$ 6363.776 \AA\ line observed for three stars of our CHEPS sample.}
\label{_A6363}
\end{figure}

\begin{figure}
\centering
\includegraphics[width=0.95\linewidth]{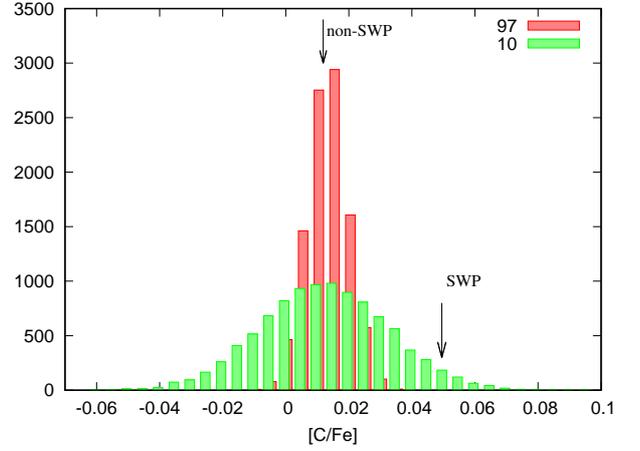}
\caption{The distributions are mean [C/Fe] for SWP(N=10) and non-SWP(N=97) points for 10,000 simulations.
Arrows mark our mean results for SWP and non-SWP from observations, respectively.}
\label{_AMC}
\end{figure}

\begin{figure}
\centering
\includegraphics[width=\linewidth]{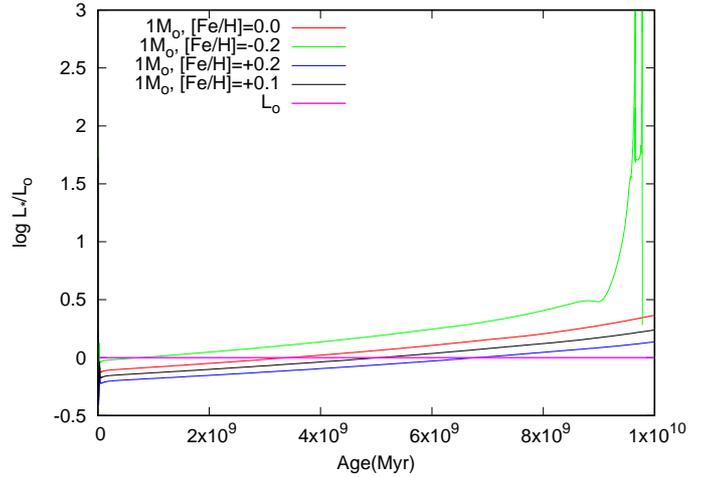}
\caption{Comparison of the theoretical evolutionary luminosity tracks 
of stars of the one solar mass of different [Fe/H] =-0.2, 0.0, +0.1 and +0.2}.
\label{_Aagesmu}
\end{figure}

\begin{figure}
\centering
\includegraphics[width=\linewidth]{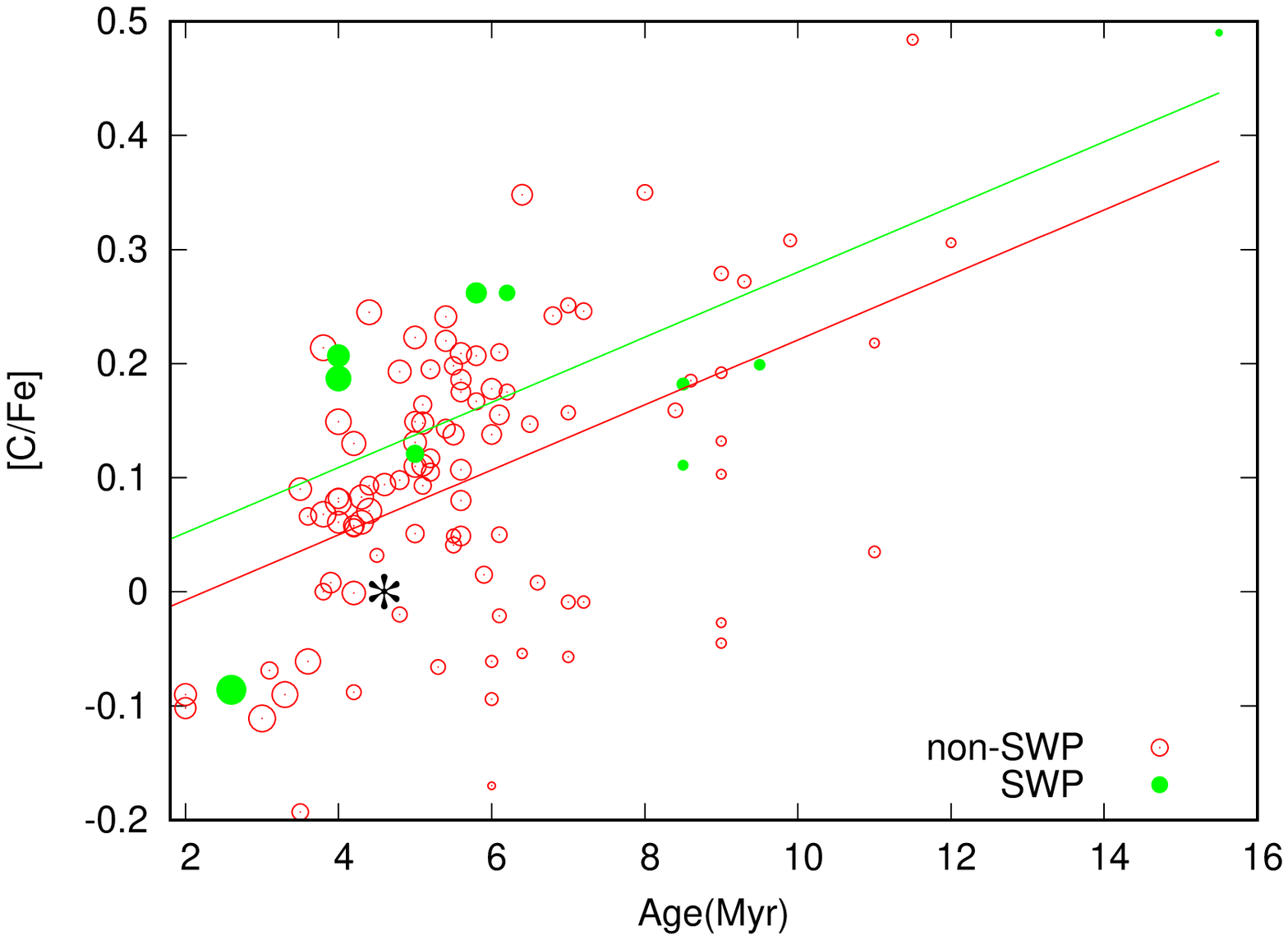}
\caption{[C/Fe] versus age of CHEPS stars. The stars with and without planets are shown by green and red colors, respectively. 
Symbols sizes are proportional to the stellar masses. The asterisk marks the Sun. }
\label{_AaC}.
\end{figure}

\begin{figure}
\centering
\includegraphics[width=\linewidth]{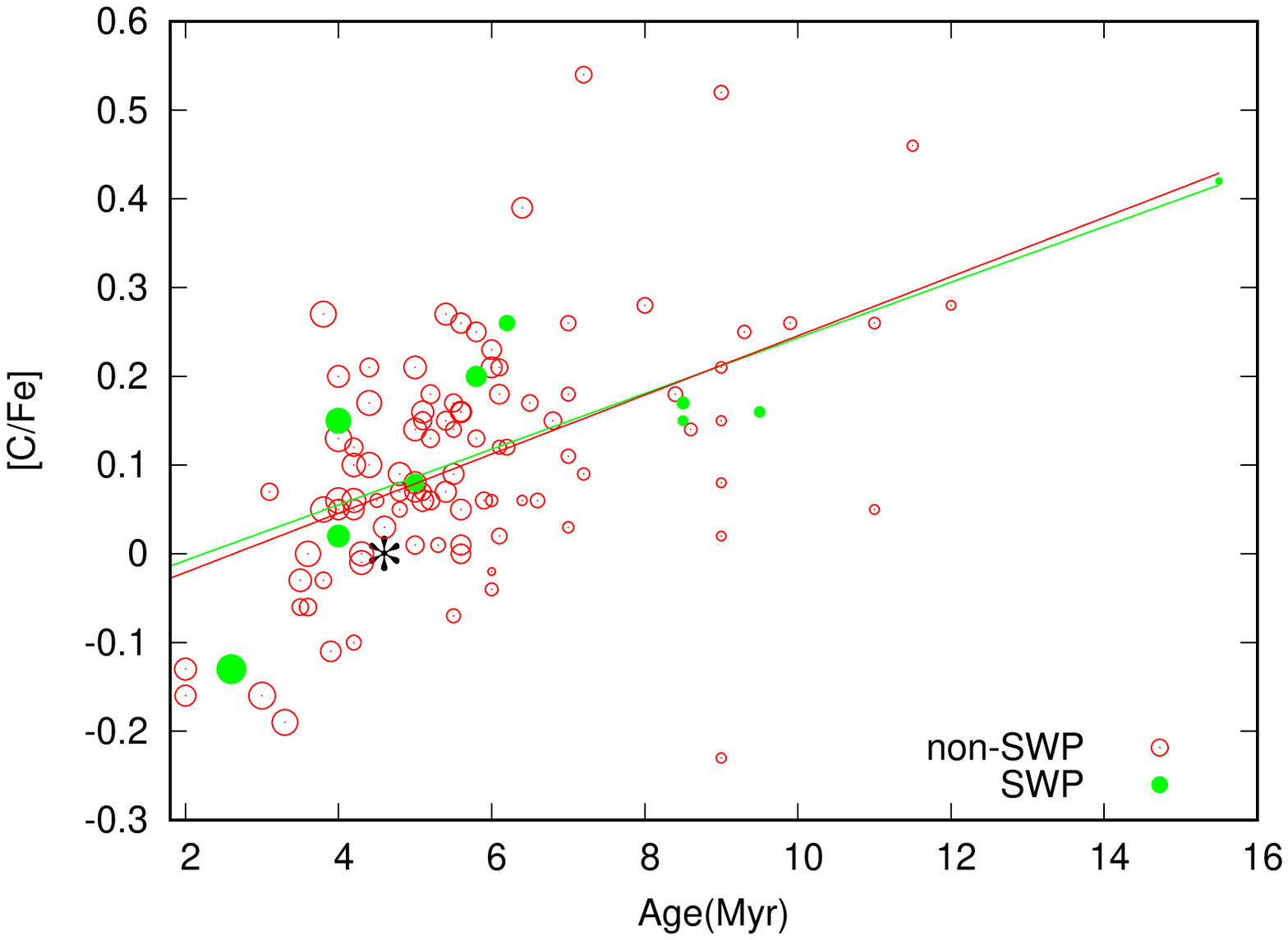}
\caption{[O/Fe] versus age of CHEPS stars. The stars with and without planets are shown by green and red colors, respectively. 
Symbols sizes are proportional to the stellar masses. The asterisk marks the Sun. }
\label{_AaO}
\end{figure}

\begin{table*}
\caption{Abundances of carbon in solar atmosphere, determined from a selected lines of the CCS}
\label{_Accs}
\begin{tabular}{cccccccc}
\hline\hline
  N  &         &               & atom,     & log N(C)&  E$_{low}$(eV)  & \Vt(\kmps) &  \Vm(\kmps) \\
     &$\lambda_1$ & $\lambda_2$ & molecule &         &                 &           &         \\
     \hline
   1 & 4218.64 & 4218.81 &  CH &  -3.68 &  0.34 &  0.85 &  3.80 \\
   2 & 4248.65 & 4248.80 &  CH &  -3.58 &  0.40 &  0.85 &  4.80 \\
   3 & 4248.86 & 4249.01 &  CH &  -3.60 &  0.40 &  0.85 &  4.60 \\
   4 & 4252.94 & 4253.06 &  CH &  -3.63 &  0.61 &  0.85 &  4.80 \\
   5 & 4253.15 & 4253.28 &  CH &  -3.64 &  0.59 &  0.85 &  4.20 \\
   6 & 4255.17 & 4255.33 &  CH &  -3.64 &  0.40 &  0.85 &  4.00 \\
   7 & 4263.90 & 4264.04 &  CH &  -3.70 &  0.62 &  0.85 &  3.60 \\
   8 & 4274.11 & 4274.27 &  CH &  -3.64 &  0.40 &  0.85 &  4.00 \\
   9 & 4356.29 & 4356.43 &  CH &  -3.54 &  0.53 &  0.85 &  4.80 \\
  10 &  4932.02& 4932.15 & C {\sc I} & -3.54  & 0.72  & 0.85  & 4.80  \\
  11 &  4992.22& 4992.38 & \ctwo  & -3.57  & 0.91  & 0.85  & 4.80  \\
  12 &  5033.68& 5033.85 & \ctwo  & -3.56  & 0.94  & 0.85  & 4.80  \\
  13 &  5052.05& 5052.25 & C {\sc I} & -3.51  & 0.77  & 0.85  & 4.80  \\
  14 &  5052.54& 5052.71 & \ctwo  & -3.56  & 0.87  & 0.85  & 4.80  \\
  15 &  5073.36& 5073.52 & \ctwo  & -3.57  & 0.86  & 0.85  & 4.40  \\
  16 &  5073.53& 5073.64 & \ctwo  & -3.58  & 0.92  & 0.85  & 3.40  \\
  17 &  5086.32& 5086.47 & \ctwo  & -3.55  & 0.92  & 0.85  & 4.80  \\
  18 &  5103.63& 5103.84 & \ctwo  & -3.58  & 0.88  & 0.85  & 4.60  \\
  19 &  5109.03& 5109.21 & \ctwo  & -3.57  & 0.88  & 0.85  & 4.60  \\
  20 &  5109.24& 5109.36 & \ctwo  & -3.57  & 0.92  & 0.85  & 3.60  \\
  21 &  5135.48& 5135.63 & \ctwo  & -3.60  & 0.89  & 0.85  & 3.80  \\
  22 &  5135.64& 5135.76 & \ctwo  & -3.59  & 0.92  & 0.85  & 3.20  \\
  23 &  5143.25& 5143.40 & \ctwo  & -3.68  & 0.94  & 0.85  & 3.80  \\
  24 &  5380.22& 5380.42 & C {\sc I} & -3.49  & 0.86  & 0.85  & 4.80  \\
\end{tabular}
\end{table*}


\end{appendix}
\end{document}